\documentclass{optica-article}

\journal{opticajournal} 

\articletype{Research Article}
\usepackage{bbold}
\usepackage{soul}
\usepackage{braket}
\usepackage{amsmath}

\begin{document}

\title{Real-time Monitoring of Multimode Squeezing}

\author{Mahmoud Kalash\authormark{1,2,*}, Aditya Sudharsanam\authormark{1,2}, M. H. M. Passos\authormark{1,2}, Valentina Parigi\authormark{3}, and Maria Chekhova\authormark{1,2}}

\address{\authormark{1}Max Planck Institute for the Science of Light, Staudtstraße 2, 91058 Erlangen, Germany\\
\authormark{2}Friedrich-Alexander Universität Erlangen-Nürnberg, Staudtstraße 7/B2, 91058 Erlangen, Germany\\
\authormark{3}Laboratoire Kastler Brossel, Sorbonne Université, ENS-Université PSL, CNRS, Collège de France, 4 place Jussieu, 75252 Paris, France\\}
\email{\authormark{*}mahmoud.kalash@mpl.mpg.de} 


\begin{abstract*} 
Multimode squeezed light is a key resource for high-dimensional quantum technologies, enhancing metrological sensitivity, boosting communication security, and enabling parallel processing in computation. Its practical potential, however, remains constrained by the inherent single-mode operation of homodyne detection, necessitating post-processing for multimode characterization. Here, we overcome this long-standing challenge by employing multimode optical parametric amplification (MOPA), enabling loss-tolerant direct detection of squeezing in each mode,  
which in turn permits mode sorting after amplification. As a result, we demonstrate, for the first time to the best of our knowledge, the real-time monitoring of multimode squeezing. With a spatial light modulator sorting the modes, we simultaneously measure squeezing in nine spatial modes co-propagating within one beam. 
Although mode sorting and filtering reduce the detection efficiency to less than $0.3\%$, we observe high-purity squeezing of up to $-7.9 \pm 0.6$ dB -- to the best of our knowledge, the highest squeezing recorded for pulsed light. Furthermore, we demonstrate real-time, loss-tolerant characterization of  continuous-variable entanglement and extend it to the detection of cluster states. Similar methods can be applied in the frequency domain, facilitating a crucial capability for scalable quantum technologies.
\end{abstract*}

\section{Introduction}
Optical quantum information science promises significant advancements in photonic technologies and heralds a new era of applications. 
Among quantum photonic resources, squeezed light stands out as the most fundamental and relatively the easiest to generate while offering a wide range of capabilities \cite{Andersen2016Apr,Schnabel2017Apr}. Even more important is multimode squeezed light, as it extends these capabilities to high-dimensional scenarios \cite{Chen2014Mar,Cai2017Jun}, making them viable for real-life applications. These include sub-shot-noise sensing and imaging~\cite{Giovannetti2004Nov,Barsotti2018Dec,Brida2010Feb,Matsuzaki2022Feb,Samantaray2017Jan}, secure communication through continuous-variable (CV) quantum key distribution~\cite{Hillery2000Jan}, boson sampling~\cite{Zhong2020Dec,Zhong2021Oct}, and  quantum computing via cluster states \cite{Larsen2019Oct,Menicucci2006Sep}.

Despite these advantages, high-dimensional scenarios have been handled inefficiently so far, limiting the full potential of multimode squeezed states. This is mainly because homodyne detection (HD) \cite{Smithey1993Mar}, the standard CV technique for characterizing quantum states, is inherently limited to single-mode operation. Specifically, HD employs a local oscillator (LO), which defines the detected mode, restricting the simultaneous retrieval of information across multiple modes. Typically, spectral and spatial modes are addressed one by one by shaping the LO~\cite{Pinel2012Feb,Cai2017Jun,Heinze2022Feb}.
Efforts to circumvent this limitation have taken several approaches. In boson sampling experiments, multiple separate squeezed light sources were used along with a corresponding number of homodyne detectors  \cite{Zhong2020Dec} — a resource-demanding approach. Alternatively, temporal modes have been employed along with delay loops and time de-multiplexers for the creation of temporal cluster states \cite{Yokoyama2013Dec,Larsen2019Oct}. Some works used array detectors to address multiple modes simultaneously \cite{Beck2000Jun,Janousek2009Jul,Armstrong2012Aug, Cai2021May}, but the information was retrieved only through post-processing. As these methods are all based on HD, they also share additional limitations. The results are highly susceptible to detection inefficiency, an issue particularly critical for array detectors. Moreover, the detection optical bandwidth is limited to that of the LO, posing challenges when dealing with states more broadband than the LO. Finally, the sampling rate is limited by the electronic bandwidth of HD.

Optical parametric amplification  
overcomes the technical limitations imposed by HD \cite{Shaked2018Feb,Frascella2019Sep,Takanashi2020Nov,Nehra2022Sep,Kalash2023Sep,Barakat2025Feb}. By sufficiently amplifying a certain quadrature $x^{(\phi)}$ while de-amplifying the conjugate quadrature, a phase-sensitive optical parametric amplifier (OPA) maps the variance of $x^{(\phi)}$ onto the output mean intensity $I^{(\phi)}$,
\begin{equation}
    I^{(\phi)}\propto e^{2G}\text{Var}\left(x^{(\phi)}\right),\label{parhomo}
\end{equation} where $G$ is the amplification gain. To infer the scaling factor between the output intensity and $\text{Var}\left(x^{(\phi)}\right) $, one measures the OPA output for the case of vacuum at its input (calibration procedure). Given sufficiently strong amplification, the detection becomes tolerant to practically any post-amplification loss as the latter simply scales down the output intensity. In addition, OPAs are typically broadband, which makes them compatible with wide-spectrum sources. Indeed, since squeezed light is often generated using OPAs \cite{Andersen2016Apr} and shares their spectral properties, it is naturally well-suited to be detected by OPAs. Finally, with direct detection  used after the OPA, the electronic bandwidth of HD is no longer a limiting factor.

Most importantly, OPAs are usually spatially and spectrally multimode \cite{Lvovsky2007Mar,Beltran2017Mar,Frascella2019Oct,Sevilla-Gutierrez2024Feb}. A multimode optical parametric amplifier (MOPA) thus has the unique capability of  amplifying multiple modes at once \cite{Barakat2025Feb}. Its modal content, determined by the nonlinear phase matching and the pump spectrum, can be engineered \cite{Li2023Jan,Patera2012Sep} to match any set of input modes, including complex modes such as orbital angular momentum ones \cite{Beltran2017Mar}. 

Here, we employ MOPA to monitor, for the first time to the best of our knowledge, multimode squeezed light in real time. By spatially expanding the MOPA pump, we engineer the shapes of its spatial modes to match those of the input quantum state. Tolerance of MOPA-based detection to loss allows us to sort the output spatial modes -- a procedure that is inevitably lossy. In particular, we measure high-purity squeezing simultaneously in nine spatial co-propagating modes within one beam. We characterize CV entanglement of many pairs of spatial modes. Moreover, we consider possible cluster states formed by various mode combinations and evaluate their quality. All these procedures are carried out in real time. The simultaneous access to multiple squeezed modes unlocks the full potential of high-dimensional quantum applications, especially ones  requiring dynamic control and feedback, including quantum metrology, communication, and continuous-variable quantum computation.
\begin{figure}[t!]
\centering
\includegraphics[width=1\linewidth]{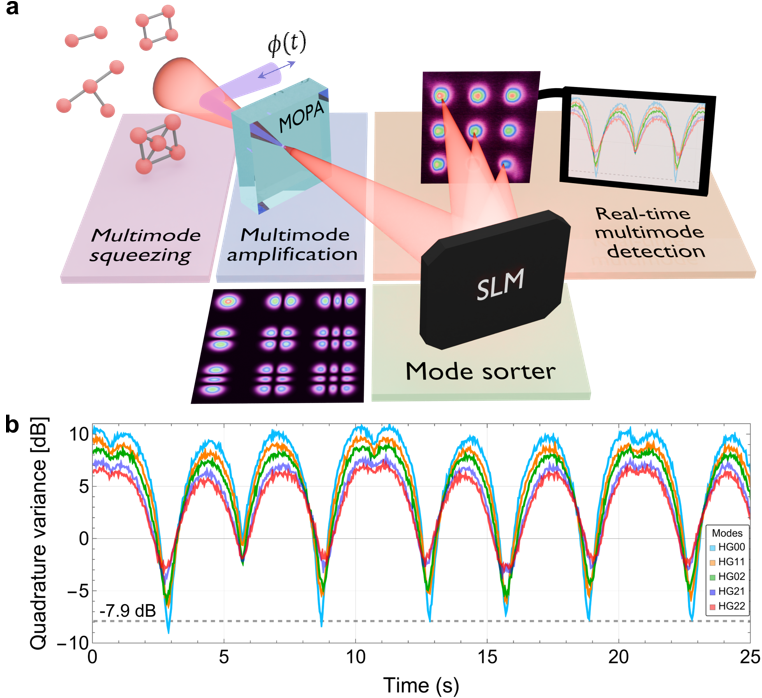} 
     \caption{(a) Generalized setup for real-time detection of spatially multimode squeezed light: after its amplification with a multimode optical parametric amplifier (MOPA), modes are sorted using a spatial light modulator (SLM). By measuring their intensities with a camera, we infer the variances of the input mode quadratures. The choice of the quadrature is determined by the pump phase. (b)  
     Traces of quadrature variances for five squeezed Hermite-Gauss modes, obtained in real-time} by scanning the pump phase.
    \label{idea}
\end{figure}

\section{Results}
Figure~\ref{idea}(a) presents the generalized setup for the detection of multimode squeezing, here applied to spatial modes. The multimode squeezed light to be monitored is injected into a MOPA with modes tailored to match the input, which is achieved by shaping the pump. In the simplest case, depending on the pump phase, either the squeezed or anti-squeezed quadratures are  amplified for all modes. Alternatively, by engineering the pump field, different quadratures can be addressed for different modes, enabling the detection of more complex states. Afterwards, a mode sorter, in this case a spatial light modulator (SLM), de-multiplexes  different modes or mode superpositions, which are then monitored through an array detector (see Methods for details). Importantly, after the amplification, the detection becomes tolerant to losses, in particular those accompanying sorting, and noise. Finally, if the MOPA amplifies quadrature $x^{(\phi)}$ of mode $(m,n)$, the degree of squeezing or antisqueezing  for this quadrature is found as $10\text{Log}\left(\frac{I_{m,n}^{(\phi)}}{I_{m,n}^{\text{vac}}}\right)$,
 where $I_{m,n}^{(\phi)}$ is the mean intensity of the mode after amplification and 
$I_{m,n}^{\text{vac}}$ 
is the mean intensity of the corresponding amplified vacuum mode, measured with the MOPA input blocked~\cite{Sharapova2015Apr,Sharapova2020Mar,Scharwald2023Nov}. 

As an example, we examine a spatially multimode squeezed vacuum state  
occupying a set of Hermite–Gaussian (HG) modes that co-propagate within a single beam. Figure~\ref{idea}(b) shows the real-time traces of the quadrature variances  
of five exemplary modes (HG$_{00}$, HG$_{11}$, HG$_{02}$, HG$_{21}$, HG$_{22}$),  
all acquired simultaneously after the MOPA stage. 
Because in our setup the squeezed modes are imaged on the MOPA modes (see Supplementary Information), we detect squeezed or antisqueezed quadratures simultaneously for all spatial modes. As the pump phase $\phi$ is scanned with time, the quadrature that is amplified goes from the squeezed to the antisqueezed and so on. 

\subsection*{Mode matching}
We generate spatially multimode squeezed vacuum using type-I collinear frequency-degenerate parametric down-conversion (PDC) in a 3 mm nonlinear crystal pumped by picosecond pulses. Such a source can generate squeezed vacuum with various spatial mode contents, depending on the pump spatial profile. In our proof-of-principle experiment, we choose the simplest Gaussian pump. The resulting set of spatial modes is close to Hermite-Gaussian beams~\cite{Fedorov2007,Miatto2012,Averchenko2020Nov}, with the size determined by the crystal length and the pump beam waist. Other modal contents can be generated by shaping the pump differently~\cite{Aadhi2017Mar}. By focusing the pump into a waist of $97\pm2$ µm inside the crystal, we obtain an effective number of 53 spatial modes, the strongest (Gaussian) mode having a waist of 23 $\mu m$. We set the squeezing parameter for the collinear emission to $G_{sq}=1.05 \pm 0.2$, corresponding to approximately 9 dB of squeezing. This value, measured without mode selection and thus reflecting the collective contribution of all spatial modes, is used as a reference to set the squeezing level of the individual modes. 
Finally, we image the squeezed vacuum modes on the MOPA. See the Supplementary Information for more details on the modal content of the squeezed vacuum.

For the MOPA, we employ the same crystal but stronger pumping (gain $G=4.4\pm0.3$ for collinear emission) to make Eq.~(\ref{parhomo}) applicable and achieve an acceptable signal-to-noise ratio while amplifying the squeezed quadrature \cite{Sharapova2020Mar}.
We characterize the MOPA modal content experimentally, by analyzing the spatial intensity correlations at its output~\cite{Averchenko2020Nov} (Methods). Because the MOPA has a higher gain than the squeezer, the angular sizes of MOPA spatial modes are larger than those of the squeezed vacuum modes, although they still form a Hermite-Gaussian set~\cite{Sharapova2020Mar}. Therefore, directly imaging the squeezed modes into the MOPA  leads to mode mismatch,  
which degrades the measurement and manifests as effective optical loss. To fix this problem, here we adjust the sizes of the MOPA modes by simply resizing the MOPA pump. In more general cases, tailoring the pump profile would be required to match the input mode structures.

\begin{figure}[t!]
\centering
\includegraphics[width=1\linewidth, keepaspectratio]{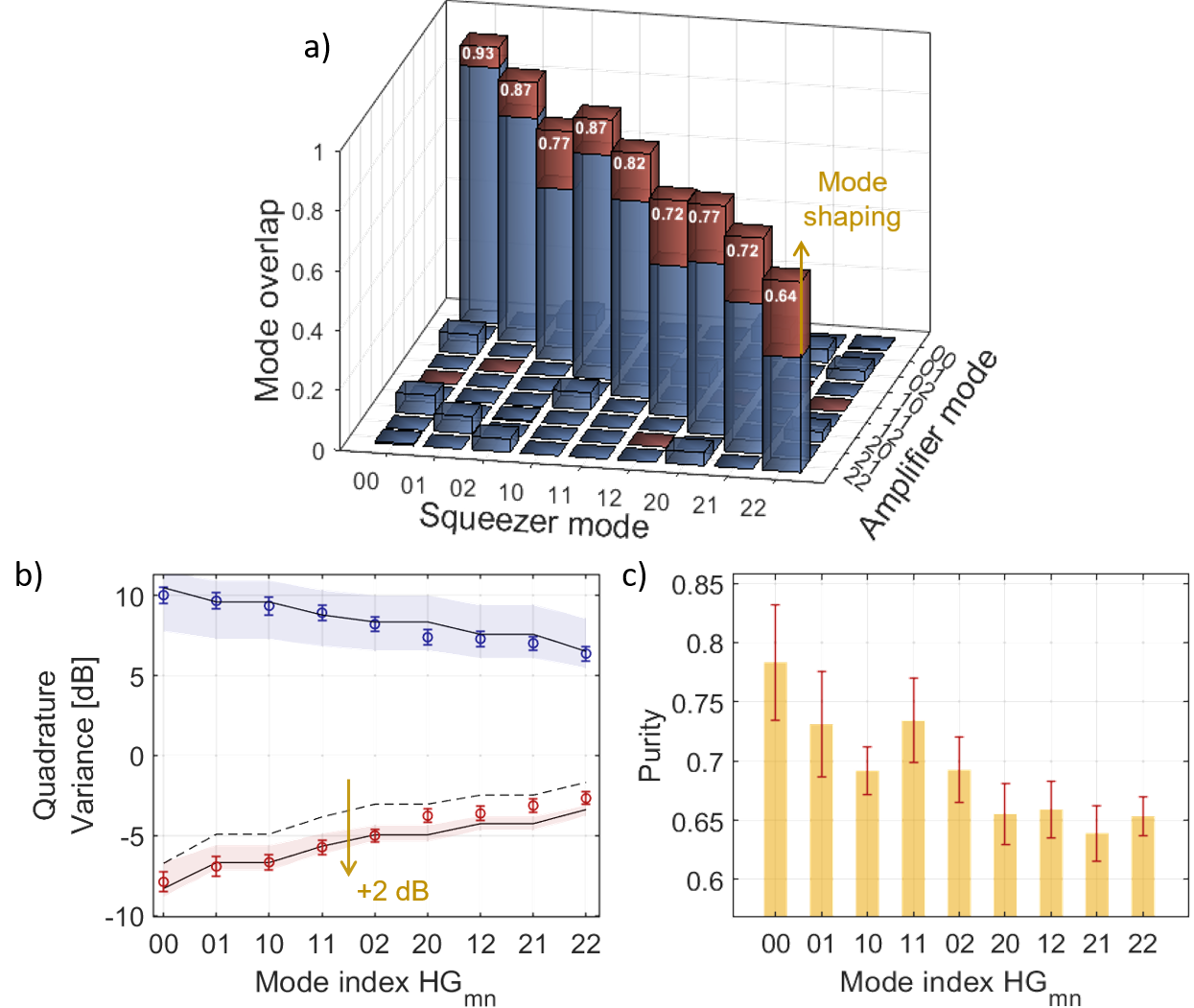}
     \caption{(a) The overlap of the MOPA modes with the modes of the squeezer without (blue bars) and with (red bars) of the MOPA pump waist opimization. (b) The antisqueezing (blue) and squeezing (red) measured for nine strongest spatial modes. Solid line shows the result of the theoretical calculation and dashed line, the result obtained without mode matching. (c) Quantum state purities obtained in experiment for all nine modes.}
    \label{characterization}
\end{figure}

In the experiment, we increase the MOPA pump waist to 145 $\mu m$, 1.5 times larger than the squeezer’s pump. This adjustment  partially compensates for the angular broadening caused by the higher gain, thus improving the mode matching. 
Figure \ref{characterization}(a) shows the calculated overlap 
(see Methods) between modes HG$_{mn}$ of the squeezer and modes HG$_{kl}$ of the MOPA. The red bars show the improvement in mode matching  
resulting from the optimized pump beam waist in the amplifier. The achieved overlap is more than 85\% for HG$_{00,10,01}$ and more than 70\% for HG$_{11,12,20,02,21}$. Mode HG$_{22}$ shows the most pronounced improvement, with its overlap increasing from 38\% to 64\% as a result of pump shaping.



 \subsection*{Real-time measurement of squeezing over multiple modes}
 The SLM sorts the amplified modes (see Methods) by converting each mode into a Gaussian-like spot at a certain location on a camera (Fig.~\ref{idea}(a)), enabling access  to several modes simultaneously. 
 The relative intensities  in these modes are measured by integrating over square areas at the centers of the spots. 
From these intensities we infer the quadrature variances and measure them while scanning the MOPA pump phase, as shown in Fig.~\ref{idea}(b).

Figure \ref{characterization}(b) shows the squeezing and antisqueezing measured for all nine modes (points). Theoretically calculated values are plotted by solid lines, the shaded area showing the uncertainty caused by the gain measurement of the squeezer. The calculation takes into account the imperfect mode overlap (see the Supplementary Information). 
 The highest retrieved degrees of squeezing and anti-squeezing are -7.9 $\pm$ 0.6 dB and 10 $\pm$ 0.5 dB, respectively, and are observed for the fundamental mode. For higher-order modes, they are gradually reducing, which matches with the expectations for our source: the higher the mode order $m+n$, the lower the squeezing. The  dashed line shows the level of squeezing that would be obtained without mode matching. We see that by increasing the pump beam waist in the MOPA, we gain about 2dB of squeezing in all modes.
 
Finally, Fig.~\ref{characterization}(c) shows the purity of the measured squeezed modes. Despite the huge detection losses (more than $99\%$ mainly due to the spatial mode sorting (see the Supplementary Information)), the purity is between 63\% and 78\%. Such a high purity is due to the parametric amplification of the state before detection.

 \subsection*{Efficient detection of cluster states}
 Having access to individual modes simultaneously, we can now build and characterize cluster states, a fundamental resource in continuous-variable quantum information \cite{Menicucci2006Sep}. These states can be generated in superpositions of squeezed-vacuum modes, forming optical nodes that are interconnected and share entanglement. A key quantity used to characterize the correlations between these cluster nodes is the nullifier $\delta$. It is evaluated based on the quadratures $X_i$ and $P_i$ of the cluster nodes as $\delta_i=\frac{P_i-\sum_j^{h_i}X_j}{\sqrt{1+h_i}}$, where $h_i$ denotes the number of neighboring nodes. Sub-shot-noise squeezing of $\delta_i$ indicates quantum correlations between the nodes of the cluster. Rather than measuring the quadratures of individual nodes separately, it is sufficient to measure the nullifier modes \cite{Cai2017Jun}. In HD, this is done by shaping the local oscillator into a superposition of the squeezing modes, to match the modes of the nullifiers.
 A full characterization of a cluster state thus requires addressing all nullifiers individually, a task that becomes increasingly demanding as the cluster size grows~\cite{Larsen2019Oct,Roh2025Feb}, requiring significant time and effort. To overcome this challenge, we propose MOPA as an efficient approach for cluster state detection. By shaping the modes of the amplifier to match those of the nullifiers, all cluster links can be characterized and monitored in real time.

 \subsection*{Real-time detection of two-node clusters}
\begin{figure}[h] 
\centering
 \includegraphics[width=1.0\linewidth]{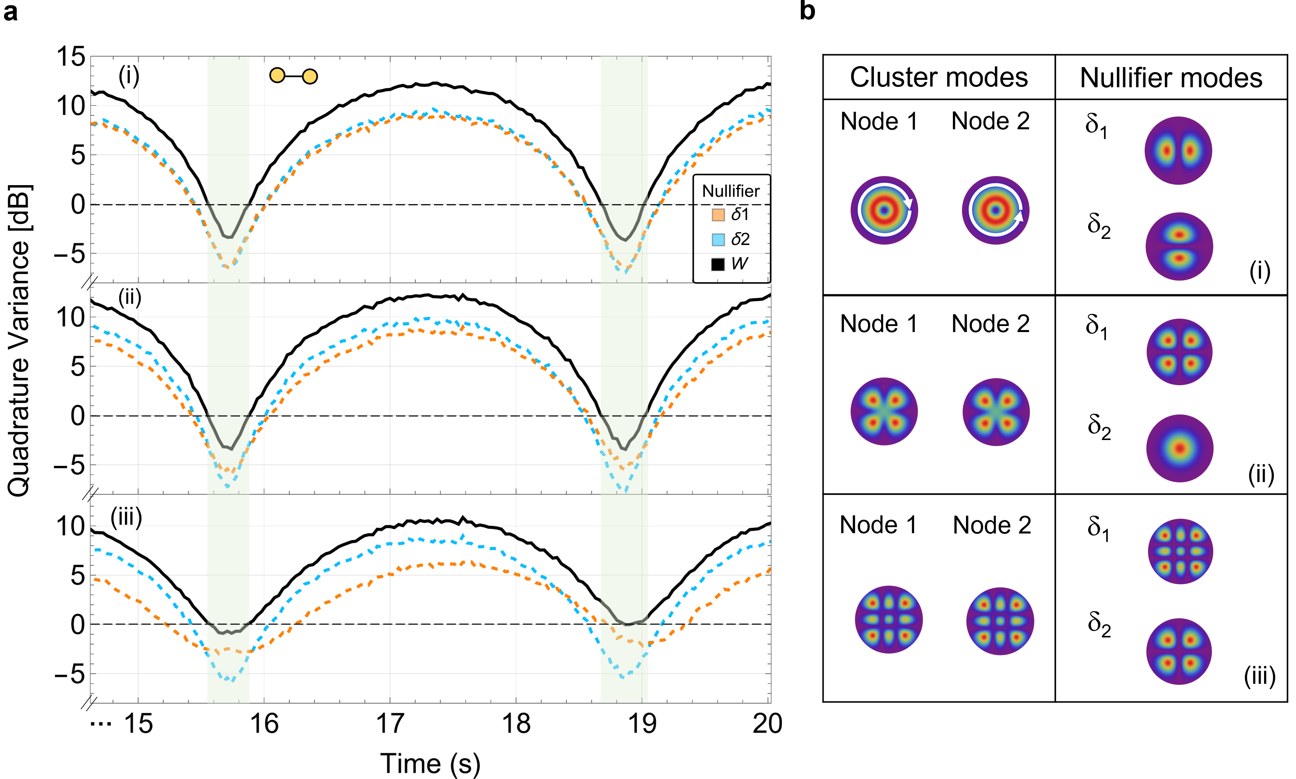}
     \caption{Simultaneously  monitoring nullifiers of two-node cluster states in real time. (a) Real-time traces of the quadrature variances for nullifiers $\delta_1$ (dashed blue) and $\delta_2$ (dashed orange) of two-node cluster states, as well as the entanglement witnesses {\textit{W}}$\equiv$ Var($\delta_1$) + Var($\delta_2$)  (solid black) as the MOPA phase is scanned. The shaded areas highlight regions where the entanglement between the two nodes is observed. (b) Cluster nodes (left) and their corresponding nullifier modes (right): HG$_{01}$ and HG$_{10}$ (i); HG$_{11}$ and HG$_{00}$ (ii), HG$_{22}$ and HG$_{11}$ (iii).} 
    \label{musq}
\end{figure}

The  simultaneous measurement of squeezing for multiple modes immediately provides a tool to characterize the entanglement of their superpositions. The first example is Laguerre-Gauss modes LG$_{1,0}$ and LG$_{-1,0}$, which are superpositions of squeezed modes HG$_{10}$ and HG$_{01}$:$\hbox{LG}_{1,0}=(\hbox{HG}_{01}+i\hbox{HG}_{10})/\sqrt{2}$ and $\hbox{LG}_{-1,0}=(i\hbox{HG}_{10}-\hbox{HG}_{01})/\sqrt{2}$. These modes form the simplest cluster state - a two-node cluster, featuring EPR-like entanglement. In this case, each nullifier variance is given by the quadrature variance of one contributing squeezed mode, and so can be fully characterized by simply monitoring these modes, $\hbox{HG}_{01}$ and $\hbox{HG}_{10}$  (see Methods). The blue and orange dashed lines in Fig.~\ref{musq}(a) show the variances of these nullifiers (i), as well as the nullifier variances for similar two-node cluster states whose nodes are superpositions of $\hbox{HG}_{00}$ and $\hbox{HG}_{11}$ (ii) and  $\hbox{HG}_{11}$ and $\hbox{HG}_{22}$ (iii). The variances are normalized to the corresponding shot-noise levels. All nullifiers are measured in real time as the phase of the pump is scanned. We see that for certain phases, all nullifiers are squeezed well below the shot-noise level.

 The difference between a two-node cluster and a CV EPR state is that for the latter, the $x$-quadrature difference and the $p$-quadrature sum for two modes are squeezed simultaneously~\cite{DuanPRL2000}, while for the former, squeezed is a nullifier that is a combination of $p$ and $x$ quadratures for different modes. A witness of entanglement for modes $1,2$ of a cluster is the sub-shot-noise of $W\equiv\text{Var}(\delta_1)+\text{Var}(\delta_2)$~ \cite{AkiraPRA2008,Takeda2019May}. Black solid lines in Fig.~\ref{musq} show this witness for three EPR-like cluster states. Green shaded areas indicate the region of entanglement, the latter being simultaneously monitored for all three clusters. In addition, owing to the real-time access to nine modes, we simultaneously monitored the remaining 33 two-node cluster states, demonstrating the parallel detection capability of our approach (see the Supplementary Information).

\subsection*{Larger cluster states} 

\begin{figure}[h]
\centering
\includegraphics[width=1\linewidth]{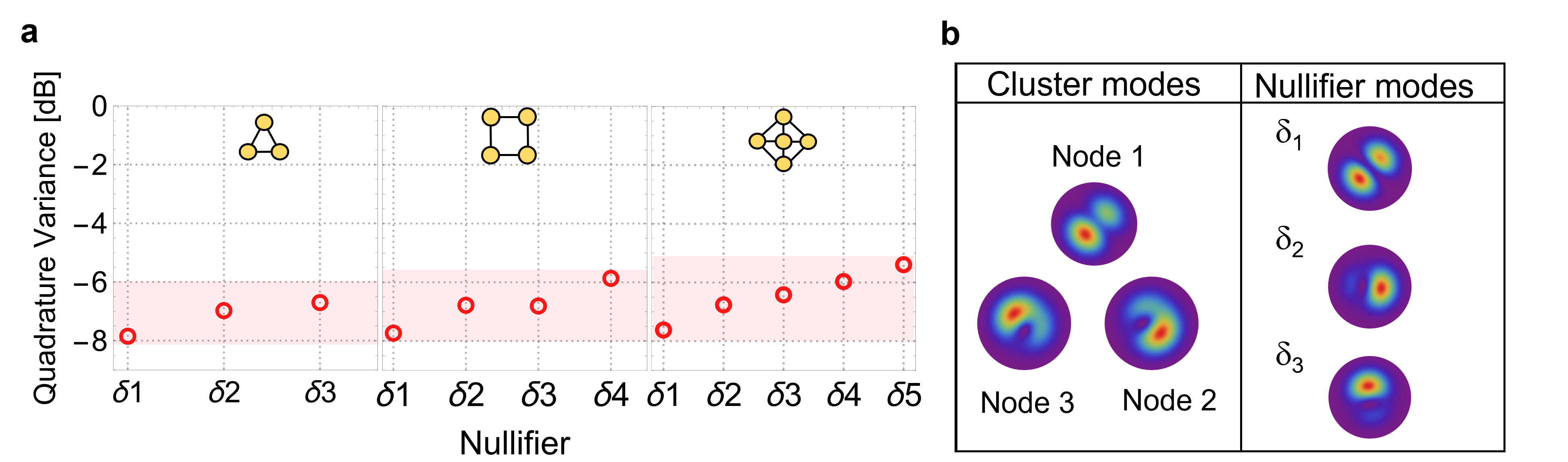}
     \caption{Two-dimensional cluster states of more than two nodes. (a) Nullifier variances of  3-, 4-, and 5-node clusters calculated from the measured variances of the contributing squeezed modes. (b) Spatial intensity profiles of node (left) and nullifier (right) modes for the three-node cluster. In this case, the involved squeezed modes are HG$_{00}$, HG$_{01}$, and HG$_{10}$. } 
    \label{cluster}
\end{figure}

Consider now clusters containing more than two nodes. Figure~\ref{cluster}(a) shows some examples: a three-node, a four-node, and a five-node  clusters formed by superpositions of HG modes. As an example,  Fig.~\ref{cluster}(b), left panel, shows the nodes of the three-node cluster. Their nullifiers (Fig.~\ref{cluster}(b), right panel) have contributions from different modes, and thus their variances cannot be measured directly without specially shaping the modes of the MOPA. However, we can calculate the squeezing of each nullifier from the measured quadrature squeezing of these contributing modes. As Fig.~\ref{cluster}(a) shows, we expect rather high degrees of squeezing: from $-7.9$dB to $-5.5$dB. In our case, the contributing modes exhibit varying degrees of squeezing, leading to different levels of nullifier squeezing within a single cluster. This variation can be mitigated by appropriately selecting the contributing modes and their respective weights~\cite{Cai2017Jun} or by tailoring the source’s modal content. 
To characterize the nullifiers of three-node or more complex clusters, engineering of the MOPA modal structure is needed.
\section{Discussion}
With the help of multimode optical parametric amplification, we have demonstrated real-time monitoring of squeezing for multiple spatial modes co-propagating within a single beam. After the simultaneous amplification of all modes, losses no longer affect the measurement, enabling mode de-multiplexing - an inevitably lossy procedure. In particular, in our experiment, the efficiency of mode sorting with an SLM was $\approx 0.5 \%$.  By sorting the modes, we were able to simultaneously access individual squeezed modes. Despite less than $0.3\%$ detection efficiency, we observed  squeezing of $-7.9 \pm 0.6$ dB with a 78\% purity in the fundamental Gaussian mode, which, to the best of our knowledge, is the highest squeezing achieved for pulsed squeezed light to date. 

In the experiment, the squeezed and amplifier modes differ in size. To optimize the measurement, we have implemented mode matching between the squeezer and the MOPA, achieving more than $70\%$ overlap for the first eight modes by making the amplifier pump broader than the source pump. 
Matching for more modes can be easily provided by further expanding the pump beam for the MOPA. 

The spatial modes of a paraxial beam form a rich system, where orthogonal mode bases can be chosen in infinitely many ways. For this reason, squeezing measured for individual modes in one basis (Hermite-Gauss) opens a path for constructing numerous cluster states out of modes in other bases, for instance, Laguerre-Gauss. For two-node clusters (EPR-like states), the nullifiers are directly measurable from the squeezing measured simultaneously for HG modes. In particular, we measured real-time traces of entanglement witnesses  
for exemplary modes LG$_{1,0}$ and LG$_{-1,0}$, as well as 35 other two-mode combinations.

Furthermore, MOPA can be used for the efficient detection of larger cluster states, enabling the characterization and monitoring of all cluster links at once. For exemplary three-, four-, and five-node clusters we have estimated the expected nullifier variances from the measured squeezing degrees for the HG modes. 

We stress that although here we only characterized nine strongest modes, the effective number of spatial modes in our setup is about $50$ and can be further increased to a few hundred by using broader pump beams. This scalability paves the way for the generation of large-scale spatially multimode squeezing.

As a proof of principle, we have demonstrated the simplest case where the MOPA amplifies simultaneously either the squeezed or the anti-squeezed quadratures across different modes. By properly engineering the amplitude and phase of the amplifier pump, different quadratures can be selectively addressed across various modes simultaneously. 

Finally, a similar method can be applied to frequency modes, where sorting can be based on nonlinear frequency conversion~\cite{Eckstein2011,Ra2017}, and tolerance to losses will be highly beneficial.

Altogether, this work completes the set of capabilities needed for robust multimode squeezed light detection and significantly expands its applicability in emerging quantum technologies.

\section{Acknowledgements}
We thank Polina Sharapova, Dennis Scharwald, and Giuseppe Patera for illuminating discussions.

\section{Funding}
This project is part of the Munich Quantum Valley, which is supported by
the Bavarian state government with funds from the Hightech Agenda Bavaria.  M. V. C. acknowledges
funding from the Deutsche Forschungsgemeinschaft (grant number 499995074). This research was funded within the QuantERA II Programme (project SPARQL), which has received funding from the European Union’s
Horizon 2020 research and innovation programme under Grant Agreement No 101017733. M. K., A. S., and M. C. are part of the Max Planck School of Photonics supported by BMBF, Max Planck Society and Fraunhofer Society. M. K. was  also funded by the Deutsche Forschungsgemeinschaft  – Project-ID 429529648 – TRR 306 QuCoLiMa (”Quantum Cooperativity of Light and Matter”).


\par
\bibliography{Optica-template}

\section{Methods}
\subsection{Experimental setup}

We employ a wide-field SU(1,1) nonlinear interferometer in a folded scheme~\cite{Frascella2019Sep}; a single 3mm bismuth triborate (BiBO) crystal cut for type-I collinear degenerate phasematching is used for both the generation and amplification of multimode squeezed vacuum (see the Supplementary Information, Fig. S1). The pump is the third harmonic of an Nd-YAG laser with 355 nm wavelength, 18 ps pulse duration, 1 kHz repetition rate and a maximum pulse energy of 70 $\mu J$. When pumped in one direction (pump 1), the BiBO crystal generates multimode squeezed vacuum. The multimode radiation is then imaged back into the crystal in a one-to-one configuration using a dichroic mirror followed by a spherical mirror. Phase-sensitive amplification is achieved by introducing a second pump (pump 2) propagating in the same direction as the back-reflected radiation. Using a piezoelectric actuator, we scan the phase between pump 2 and the input signal. The power of pump 1 is set at 8 mW to achieve $G_{sq}=1.05\pm0.2$, theoretically corresponding to approximately 9 dB of squeezing in the collinear emission (in the absence of losses). Meanwhile, pump 2 power is adjusted to achieve an amplification gain of $G=4.4\pm0.3$. This satisfies Eq. (1) with a squeezing detection accuracy of $>99\%$ for the strongest nine modes (see the Supplementary Information). Additionally, it ensures an acceptable signal-to-noise ratio when amplifying the squeezed quadrature, where the output mean intensity for each mode (m,n) follows $\braket{I_{m,n}^{(\phi)}}=\sinh^2(G_{m,n}-G_{{sq}_{m,n}})$, where $G_{m,n}$ and $G_{{sq}_{m,n}}$ are the gains of the amplifier and the squeezer, respectively, for this mode.
Afterwards, a system of lenses magnifies the far field of the amplified radiation. An SLM (Hamamatsu LCOS-SLM X10468-06), located in the far field, sorts out the amplifier modes. Finally, an sCMOS camera is used to monitor in real time the intensities of the sorted modes.   
\subsection{Calculation of the modal content}
To calculate the modal content for the squeezer (where measurement of mode shapes was impossible because of the low photon flux) and the amplifier (where calculation was compared with the measurement results), we run the integro-differential equations governing high-gain collinear degenerate PDC~\cite{Sharapova2020Mar, Scharwald2023Nov}. Namely, we find numerically the gain-dependent functions $\eta(q,q',g)$ and $\beta(q,q',g)$ from the Bogoliubov transformations connecting the input and output annihilation operators $\hat{a}^{in},\hat{a}^{out}$: 
\begin{equation*}
    \hat{a}^{out}(q,g)=\int dq'\eta(q,q',g)\hat{a}^{in}(q')+\int dq'\beta(q,q',g)[\hat{a}^{in}(q')]^\dagger.
\end{equation*}
Here, $q$ is the transverse wavevector and $g$ is the squeezing parameter (amplification gain) measured in the collinear direction. Finally, by applying the joint Schmidt decomposition~\cite{Scharwald2023Nov} to $\eta(q,q',g)$ and $\beta(q,q',g)$, we find the one-dimensional modal content as
\begin{equation*}
    \begin{split}
        \beta(q,q',g)&=\sum_n \sqrt{\Lambda_n} u_n(q,g)\psi_n(q',g),
    \end{split}
\end{equation*}
where $\psi_n(q,g)$ and $u_n(q,g)$ are the input and output gain-dependent modes, respectively, and $\Lambda_n=\text{sinh}^2(g_n)$ is the mean photon number in mode $n$, which defines the squeezing (amplification gain) for this mode. To pass to 2D modes, we use the factorability of the modes in $x$ and $y$ directions: $\psi_{m,n}(q_x,q_y)=\psi_m(q_x)\psi_n(q_y)$, $u_{m,n}(q_x,q_y)=u_m(q_x)u_n(q_y)$. Meanwhile, the gain of a 2D mode (m,n) is $G_{mn}=\frac{g_mg_n}{g}$. See Supplementary Material for more detail.  

\subsection{Experimental reconstruction of the modal content}
\begin{figure}
    \centering
    \includegraphics[width=\linewidth]{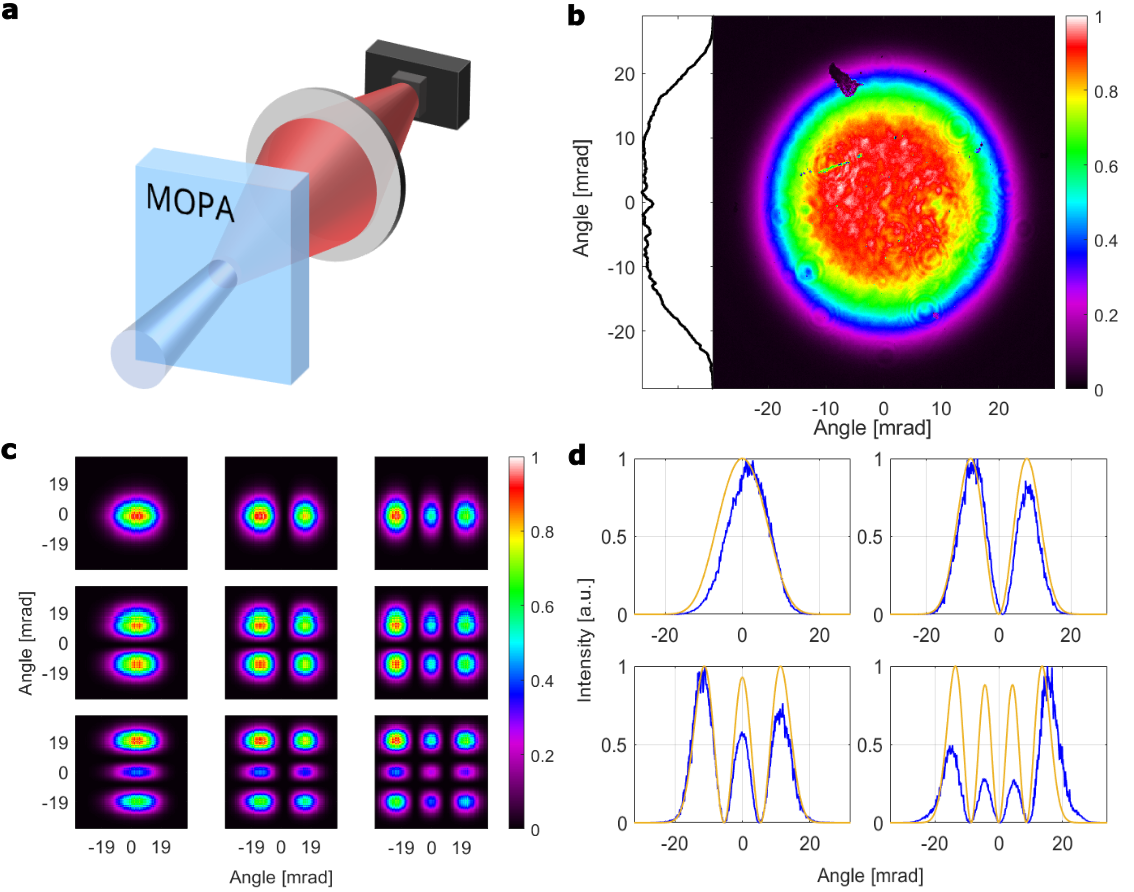}
    \caption{Reconstruction of the MOPA spatial modes. (a) the setup; (b) the far-field intensity distribution $I(\theta_x,\theta_y)$; (c) the mode shapes $|u_{m,n}(\theta_x,\theta_y)|^2$; (d) Calculated (orange) and measured (blue) one-dimensional mode shapes $|u_m(\theta_x)|^2$ for $m=1,2,3,4$.}
    \label{fig:modes}
\end{figure}

To reconstruct the MOPA mode shapes, we block the quantum state at the input, so that the MOPA amplifies only vacuum, and examine the far-field intensity distribution of the emitted radiation (Fig.~\ref{fig:modes}(a)). This distribution spans about 20 mrad FWHM and has a flat-top profile, typical for type-I collinear-degenerate phase matching~\cite{Frascella2019Sep} (Fig.~\ref{fig:modes}(b)). To retrieve the modes, we acquire 1250 single-shot intensity distributions $I(\theta_x,\theta_y)$, where the angles $\theta_x,\theta_y$ are related to the transverse wavevectors $q_x,q_y$ as $\theta_{x,y}=q_{x,y}/k$ and $k$ is the full wavevector. 

The MOPA modes form an orthogonal set with no correlations to each other, and so they are the modes with maximal achievable squeezing. As shown in Refs.~\cite{Averchenko2020Nov,Frascella2019Oct}, these modes for the bipartite (signal+idler) system coincide with the coherent modes of a single (signal or idler) subsystem. We distinguish signal and idler subsystems  by dividing the far field into the right and left, or upper and lower, parts. The coherent modes of a single subsystem can be then reconstructed from the first-order correlation function $g^{(1)}$, which, due to the thermal  statistics of a single subsystem, is related to the second-order correlation function $g^{(2)}$ via the Siegert relation  $g^{(2)}=1+|g^{(1)}|^2$. Meanwhile, $g^{(2)}(\theta_x,\theta'_x)=C(\theta_x,\theta'_x)/(\braket{I(\theta_x)}\braket{I(\theta'_x)})$+1, where $C(\theta_x,\theta'_x)$ is the  intensity covariance function
\begin{equation}
    C(\theta_x,\theta'_x)=\braket{I(\theta_x)I(\theta'_x)}-\braket{I(\theta_x)}\braket{I(\theta'_x)},
\end{equation}
calculated from the ensemble of single-shot one-dimensional intensity distributions $I(\theta_x)$ at $\theta_y$ fixed. Using all these relations, we obtain the amplifier one-dimensional spatial modes $u_m(\theta_x)$ by applying the singular-value decomposition to the measured covariance distribution as
\begin{equation}
    C(\theta_x,\theta'_x)\propto\Big[\sum_m\;\lambda_m u_m(\theta_x)u_m^*(\theta'_x)\Big]^2,\end{equation}
with $\lambda_m$ defining the weight of mode $u_m(\theta_x)$. Similarly, we obtain the one-dimensional modes for the $\theta_y$ angle. Finally, to obtain the two-dimensional modes, we 
make use of the factorability of the modes in the case of collinear degenerate PDC: $u_{m,n}(\theta_x,\theta_y)=u_m(\theta_x)u_n(\theta_y)$. 
Figure~\ref{fig:modes}(c) shows the reconstructed intensity distributions of the nine strongest modes, from HG$_{00}$ to HG$_{22}$. 

The mode matching between the output modes of the squeezer $u'_{m,n}(\theta_x,\theta_y)$ and the input modes of the amplifier $\psi''_{k,l}(\theta_x,\theta_y)$ is evaluated by the overlap integral $|\kappa_{m,n,k,l}|^2=|\int_x d\theta_x\int_y d\theta_y[\psi''_{k,l}(\theta_x,\theta_y)]^*u'_{m,n}(\theta_x,\theta_y)|^2$,  As the parametric gain increases, the far-field modes broaden~\cite{Sharapova2020Mar}. Meanwhile, as the pump waist broadens, the far-field modes get narrower~\cite{Fedorov2007}. This enables matching the modes of the squeezer and of the amplifier in our case by softer focusing the pump of the latter~\cite{Barakat2025Feb}. 

\subsection{Detectable squeezing per mode}
Imperfect mode matching has the same effect as losses on the detectable squeezing per mode. We can thus calculate the expected squeezing as 
\begin{equation*}        \text{Sq}\left(x^{(\phi)}_{m,n}\right)=10\text{Log}(\text{e}^{-2G_{sq,m,n}}|\kappa_{m,n,k,l}|^2+R_{m,n}),
\end{equation*}
where $R_{m,n}=1-|\kappa_{m,n,k,l}|^2$ are the losses accompanying the detection of an HG$_{m,n}$ mode due to imperfect mode matching.
Here, we considered the mode overlap matrix $|\kappa_{m,n,k,l}|^2$ (Fig. \ref{characterization}(a)) to be almost diagonal as the off-diagonal terms are arguably negligible. However, for a more complete measurement, the detected squeezing must be corrected to the off-diagonal contributions (see~\cite{Barakat2025Feb}).

\subsection{Mode Sorting}
A projective mode sorter is implemented using a phase-only SLM (Hamamatsu LCOS-SLM X10468-06) with a resolution of $20\, \mu$m per pixel, and computer-generated holograms (see the Supplemental Material). After the amplification, the far field of the radiation is imaged onto the SLM using a system of lenses ($f_1 = 30$ cm, $f_2  = 40$ cm, $f_3 = 20$ cm), resulting in a beam with the FWHM of the fundamental mode $1.25$ mm. The holograms are encoded onto the SLM using complex amplitude modulation with a phase function $H(r)\, =\, \exp\{i \theta(r)\}$, where $\theta(A, \Phi)$ is a function of the amplitude $A(r) \in [0,1]$ and phase $\Phi \in [-\pi, \pi]$ of the mode to be sorted. The hologram is designed to sort the mode into the first diffraction order with $\theta(A, \Phi)\, =\, f(A) \sin(\Phi)$ \cite{Arrizón2007Nov}, where $f(A)$ is obtained by inverting the first-order Bessel function in $J_1[f(A)] = 0.58A$. This technique was chosen as it sorts modes with a high signal-to-noise ratio \cite{Clark2016Mar}. A sinusoidal grating is implemented to achieve the spatial separation of the first diffraction order. Simultaneous sorting of nine modes is achieved by multiplexing their holograms, so that the hologram of each mode is overlayed with a grating of a different spatial frequency~\cite{Guzman2017Oct}. An sCMOS camera (Andor Zyla 5.5) placed in the Fourier plane of a lens ($f_4 = 30$ cm) detects the projections of nine strongest Schmidt modes present in the incident radiation. The projections form a grid in the detection plane, allowing the intensity contribution $\zeta_{m,n}$ (weight) of individual modes to be inferred by measuring  the relative brightness in the center of each sorted mode. It is calculated as
\begin{equation}\label{eq_weightslm}
    \zeta_{m,n}\, =\, \frac{I_{m,n}}{\sum_{m,\, n} I_{m,n}},
\end{equation}
where $I_{m,n}$ represents the integral intensity corresponding to the black squares in Fig. \ref{musq}(a). See the Supplemental Material for more detail.

\subsection{Cluster states}
A squeezed vacuum consisting of $N$ Schmidt modes can be described by the quadrature vector
\begin{equation}
    q^s = \{ x^s_1\;, x^s_2\;, x^s_3\;, ..., x^s_N; \; p^s_1\;, p^s_2\;,p^s_3\;, ..., p^s_N \}, 
\end{equation}
with $x^s_i$ ($p^s_i$) being the position- (momentum-) like quadrature of the $i^{th}$ mode. The quadratures of the n-node cluster state to be obtained from the multimode squeezed vacuum,
\begin{equation}
     Q^c = \{ X^c_1\;, X^c_2\;, X^c_3\;, ..., X^c_n; \; P^c_1\;, P^c_2\;,P^c_3\;, ..., P^c_n \},
\end{equation}
are found by applying a unitary transformation $U$ to the Schmidt-mode quadratures as
\begin{equation}\label{eq:ChangeBasisQuad}
   Q^c = U\; q^s = \begin{bmatrix}
a & -b \\
b & a 
\end{bmatrix} \; q^s,  
\end{equation}
where
\begin{equation}\label{eq:MatrixX} 
    a = ( \mathbb{1} + A^2 )^{-1/2},
\end{equation}
\begin{equation} \label{eq:MatrixY} 
     b =  A\; ( \mathbb{1} + A^2 )^{-1/2},
\end{equation} and $A$ is the adjacency matrix of each cluster topology. For instance, the adjacency matrices corresponding to the clusters considered in Fig.~\ref{cluster}(a) are 

\begin{equation}\label{eq.AdjM2}
A_{3} = \begin{bmatrix}
0 & 1 & 1 \\
1 & 0 & 1 \\
1 & 1 & 0 
\end{bmatrix}, A_{4} = \begin{bmatrix}
0 & 1 & 0 & 1 \\
1 & 0 & 1 & 0 \\
0 & 1 & 0 & 1 \\
1 & 0 & 1 & 0 
\end{bmatrix}, \text{and}\; A_{5} = \begin{bmatrix}
0 & 1 & 1 & 1 & 1 \\
1 & 0 & 1 & 0 & 1\\
1 & 1 & 0 & 1 & 0\\
1 & 0 & 1 & 0 & 1 \\
1 & 1 & 0 & 1 & 0
\end{bmatrix}.
\end{equation}
Finally, the normalized nullifier of node $i$ can be obtained as
\begin{equation}\label{eq:nullifier}
\delta_i = \frac{ P^c_i - \sum_{j}^{N}\; A_{ij} \; X^c_j}{\sqrt{1+h_i}},
\end{equation}
where $N$ is the total number of nodes and $h_i$ is the number of adjacent nodes to the $i^{th}$  node. 

As an example, for a 2-node cluster state, which has the following adjancency matrix
\begin{equation}
    A_{2} = \begin{bmatrix}
0 & 1 \\
1 & 0 
\end{bmatrix},
\end{equation}  
the nodes quadratures will be written as
\begin{equation}
\begin{split}
     X^c_1 &= \frac{x^s_1 - p^s_2}{\sqrt{2}}, \;\;\;\; X^c_2 = \frac{x^s_2 - p^s_1}{\sqrt{2}}\\ \\
    P^c_1 &= \frac{p^s_1 + x^s_2}{\sqrt{2}},\;\;\;\; P^c_2 = \frac{p^s_2 + x^s_1}{\sqrt{2}},\end{split}
\end{equation}
meanwhile their nullifiers and variances take the following form
\begin{equation}
\begin{split}
\delta_1 &= \frac{P^c_1 - X^c_2}{\sqrt{2}}=p^s_1, \; \delta_2 = \frac{P^c_2 - X^c_1}{\sqrt{2}} =p^s_2, \\ \\
\Delta^2\delta_1 &=\Delta^2p^s_1, \;\;\;\; \Delta^2\delta_2 =\Delta^2p^s_2. 
\end{split}
\end{equation}
Clearly, in this case, each individual nullifier is directly related to one contributing squeezed mode. Since both squeezed modes are inherently the modes of the amplifier, we can  monitor the nullifiers of this cluster without further adjusting our experimental settings. This situation is realized for three two-node clusters {(Fig.~\ref{musq})}, whose nodes are superpositions of modes HG$_{01}$ and HG$_{10}$ (i), HG$_{11}$ and HG$_{00}$ (ii), and HG$_{22}$ and HG$_{11}$ (iii).

However, for more complex cases, such as the 3-nodes cluster, calculations lead to  
\begin{equation}
\begin{split}
     \Delta^2 \delta_1 &= 0.949967\; \Delta^2 p^s_1 + 0.0250165\; \Delta^2 p^s_2 + 0.0250165\; \Delta^2 p^s_3,\\ \\
      \Delta^2 \delta_2 &= 0.0250165 \;\Delta^2 p^s_1+0.949967 \;\Delta^2 p^s_2+0.0250165 \;\Delta^2 p^s_3,\\ \\
      \Delta^2 \delta_3 &= 0.0250165 \; \Delta^2 p^s_1 + 0.0250165 \;\Delta^2 p^s_2 + 0.949967\; \Delta^2 p^s_3.
\end{split}
\end{equation}
In this case, individual nullifiers have contributions from several squeezed modes. To directly measure these nullifiers, we therefore need to engineer the modes of the amplifier.

In particular, the mode sets forming the 3-node, 4-node, and 5-node clusters in Fig.~\ref{cluster}(a) are {HG$_{00}$;HG$_{01}$;HG$_{10}$}, {HG$_{00}$;HG$_{01}$;HG$_{10}$;HG$_{11}$}, and {HG$_{00}$;HG$_{01}$;HG$_{10}$;HG$_{11}$;HG$_{02}$}, respectively.

\end{document}